# STUDY OF THE CORRELATIONS BETWEEN THE HIGHEST ENERGY COSMIC RAY SHOWERS AND GAMMA RAY BURSTS

Todor Stanev and Robert Schaefer

Bartol Research Institute, University of Delaware

Newark, DE 19716, U.S.A.

Alan Watson

Department of Physics, University of Leeds, LS29JT, Leeds, U.K.

## Abstract

We examine the correlation between the arrival direction of ultra high energy showers and the gamma ray bursts from the third BATSE catalog. We find no correlation between the two data sets. We also find no correlations between a pre-GRO burst catalog and the Haverah Park UHE shower set that cover approximately the same period of time.

## 1. Introduction

It has been observed [1] that the two highest energy cosmic ray air showers come from directions that are within the error boxes of two remarkable gamma ray bursts (GRB) detected by BATSE [4] with a delay of O(10) months after the bursts. This observation has provoked theorists [2, 3] to search for a common origin for these two phenomena, the sources of which are not known.

We attempt to follow up on this observation and look for coincidences between gamma ray bursts and a statistically larger sample of Ultra High Energy Cosmic Rays (UHECR).

## 2. Correlations with the 3B catalog

A concentration of the arrival directions of UHECR of energy exceeding $4 \times 10^{19}$ eV around the plane of weight for the cosmologically local galaxies ($z \lesssim 0.03$) (the Supergalactic Plane [5]) was recently reported for a sample of events coming from four different Northern Hemisphere detectors [6]. The sample used for this work consists of 42 events from the Haverah Park, AGASA, Volcano Ranch and Yakutsk experiments. Statistically the sample is dominated by the Haverah Park events (27). As in Ref. [6] we use only events detected at zenith angle less than 45°, for which the direction and energy is estimated with highest precision.

The 3B catalog consists of 1122 gamma ray bursts detected between 21 April 1991 and 19 Sept 1994. Not all of these are in the field of view of the four air shower arrays. The southernmost of them, Volcano Ranch, is at a latitude of 32.2°N, which means that no GRB with declination south of −9.8° is in the field of view of the UHECR detectors. There are 692 GRB at this declination in the 3B catalog. For most of the statistics, however, the field of view is determined by the Haverah Park array, which only sees declinations north of 9°.

A coincidence between the arrival directions of a GRB and UHECR was determined as an event where the angle between the directions of the two events is less than the joint error box and less than 10°. This number is arbitrary and was chosen to avoid calling 'coincidences' the small number of event pairs that are far apart, but still qualify because of the large GRB error box. The joint error box is the quadratic sum of the two error boxes, with the shower error box was taken to be 3° for all air shower arrays.

Table 1: *Results from 10,000 Monte Carlo runs used to determine the statistical significance of the coincidences between the arrival directions of GRB and UHHCR.*

| N≥ | 90 | 100 | 110 | 120 | 130 | 140 | 150 | 160 | 170 | 180 |
|---|---|---|---|---|---|---|---|---|---|---|
|  | 1.000 | 1.000 | 0.997 | 0.962 | 0.804 | 0.482 | 0.181 | 0.040 | 0.004 | 0.001 |

With a statistics of 692 GRB and 42 UHECR the total number of possible coincidences (gamma ray bursts inside the field of view for each detector × the number of showers detected) is 23,849. 144 actual coincidences were observed for the event sample. Each of the UHECR events contribute to this total number with an average number of 3.4 coincidences per air shower event. There is no trend for increase of the number of coincidences with the shower energy. If we split all showers in three groups of energy $4 - 6 \times 10^{19}$, $6 - 10 \times 10^{19}$ and $>10^{20}$ eV, the average number of coincidences per shower is 3.6, 3.3, and 3.0 respectively.

To determine the statistical significance of this result we ran a simple Monte Carlo, consisting of looking for coincidences between the 3B catalog and UHECR samples of 42 events with the same declinations as the original one, but with random right assensions. Table 1 shows the probability that more than N coincidences are found in 10,000 Monte Carlo runs.

The experimental number of 144 coincidences is almost in the peak of the distribution with a chance probability close to 0.5. No correlation is therefore found between the arrival directions of the 42 UHECR and the bursts from the 3B catalog.

The idea of correlating the 3B catalog with this data sample is in contradiction with the simplest coincidence scenario [1], in which the GRB and the UHHECR are produced (quasi) simultaneously. The cosmic ray nucleus is observed later because of the diffusive propagation in the intergalactic magnetic fields. The analysis just described requires the implicit assumption that the sources of all gamma ray bursts are repeaters.

### 3. Correlations with pre–BATSE gamma ray bursts

A limited sample exists that allows us to study the simultaneous GRB/UHECR scenario. The Haverah Park array started detecting UHE cosmic rays in July 1967 ( July 1968 with full area ) and was operated until 31 July 1987. Pre BATSE GRB catalogs exist [7] that cover a large fraction of that period ( 3 July 1969 – 13 June 1979 ). We have correlated the Klebesadel et al. [7] data set with the Haverah Park event sample above $4\times10^{19}$ eV.

Only events detected simultaneously by two or more satellites are included in this gamma ray bursts data set. There are three types of bursts in terms of location: we call *type 1* those bursts with good positional information (RA, dec and error box radius); *type 2* are bursts located in two possible positions in the sky, and for *type 3* the location is only known within an annulus on the sky.

Since coincident *type 3* bursts would contribute to the coincidence statistics with very small weight, we have used only burst *types 1 & 2* to look for coincidences. A coincident event is defined as before. We found four coincidences between the 27 Haverah Park showers of energy above $4\times10^{19}$ eV and zenith angle $\theta$ <45° and 22 gamma ray bursts of *types 1 & 2* in the Haverah Park field of view ($\delta$ >9°). Fig. 1a shows

Table 2: *Four pairs of a gamma ray burst and UHECR detected by Haverah Park, which were found to be coincident in the joint error box (see text). Columns 1 to 6 describe the GRB and 8&9 refer to the UHECR shower. The dates for both sets are given in yymmdd and $\delta\Psi$ is the solid angle distance in degrees.*

| GRB cat # | day | type | RA | dec | error | $\delta\Psi$ | E(EeV) | day |
|---|---|---|---|---|---|---|---|---|
| B 720117 | 720514 | 2 | 322.0 | 50.0 | 5.0 | 3.0 | 115.30 | 680208 |
| BS 1144+78 | 720514 | 1 | 176.0 | 78.0 | 3.0 | 1.3 | 81.10 | 720402 |
| B 730721 | 730721 | 2 | 192.0 | 65.0 | 10.0 | 6.2 | 98.20 | 800112 |
| BS 1412+79 | 790613 | 1 | 213.0 | 79.0 | 0.5 | 1.0 | 42.60 | 850731 |

the locations of all these events in polar projection. Table 2 identifies the coincident events.

10,000 Montecarlo runs as described above were used to estimate the statistical significance of the four coincidences. The number of coincidences distribution from the Montecarlo peaks at 3 and 4 coincidences. The probability that 4 coincidences are random is 0.539. No coincidences were thus found between these two data sets that were taken simultaneously over a period of more than 10 years. In two of the four coincident pairs the air shower is detected earlier than the gamma ray burst.

**4. Discussion**

We have examined the set of UHECR detected by four northern hemisphere detectors that is available to us and have found no correlations between them and and the gamma ray bursts from the 3B catalog that are not consistent with random. No correlations were found either between a pre–BATSE GRB catalog [7] and the Haverah Park shower set that were taken simultaneously from 1969 to 1979.

One has to be, however, very careful in interpreting the results of this study. It is difficult to make a statement about correlations when the exposures of the two types of detectors (GRB and UHECR) to the same regions of the sky is not known and, in the case of the pre-BATSE bursts analysis, impossible to estimate. Although we do not find any correlated pairs of events, we should not forget that the highest energy cosmic ray (detected by the Fly's Eye experiment and not present in our data set) comes from the same direction as the highest–total–fluence burst in the 3B catalog.

It is also possible that the UHECR are deflected in a systematic way, that masks the correlation with the GRB [2, 3], on propagation in the galactic and extragalactic magnetic fields. Milgrom and Usov [1] argue that the deflection angles expected for the highest energy events ($E \gtrsim 2 \times 10^{20}$ eV) are uncertain but small, possibly within the joint error boxes. To have a reasonable statistics for a correlation study, we go down in energy here by a factor of 5, which would correspondingly increase the systematic error box, which is due to magnetic deflection, for the UHECR.

The absence of correlation between the arrival directions of UHECR and GRB does not rule out model for common cosmological origin of the phenomena [2, 3], which would both have very small anisotropy. GRB distribution has been shown to be isotropic, while there is evidence for a correlation of UHECR with the large scale structure of cosmologically nearby galaxies [6]. The models of cosmological origin of UHECR would be in contradiction with data if this correlation is confirmed with higher experimental statistics.

**Acknowledgements.** The work of T.S. is supported in part by the U.S. Department of Energy.

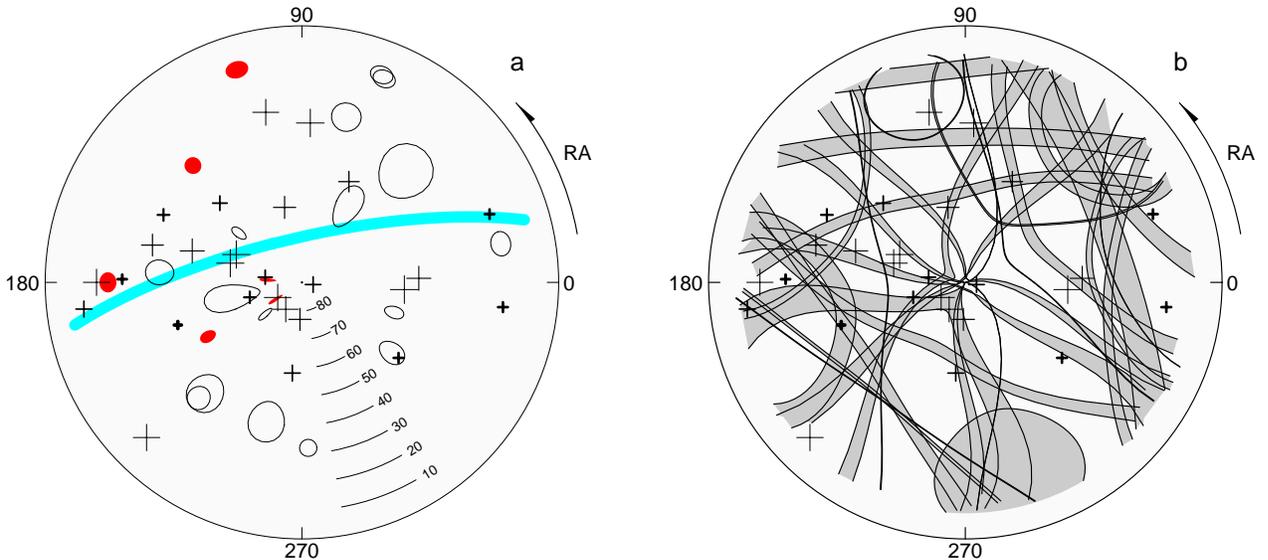

Fig. 1 *(a)* Locations of the gamma ray bursts of *type 1 & 2* and of the Haverah Park UHECR of energy above $4 \times 10^{19}$ eV. GRBs of *type 1* are shown with their error boxes as shaded areas and those of *type 2* as open circles in polar coordinates. Haverah Park events are shown with crosses. The size of the crosses is in reverse proportion to and the width of the line is proportional to the event energy. The shaded band shows the position of the Supergalactic plane. *(b)* Same for GRBs of *type 3*. Single lines correspond to burst located in a very thin annulus.